\documentclass[aps,prd,nofootinbib,amsmath,amssymb,superscriptaddress,twocolumn,10pt]{revtex4}

\usepackage{graphicx}
\usepackage{dcolumn}
\usepackage{bm}
\usepackage{amssymb}
\usepackage{latexsym}
\usepackage{booktabs}
\usepackage{amsmath}
\usepackage{multirow}
\usepackage[colorlinks=true, linkcolor=red, citecolor=blue]{hyperref}

\def\be{\begin{equation}}
\def\ee{\end{equation}}
\def\ba{\begin{eqnarray}}
\def\ea{\end{eqnarray}}

\newcommand{\ck}{c_K}
\newcommand{\tppi}{\ck p_T \Pi_T}

\begin{document}

\title{Exploring interacting holographic dark energy in a perturbed universe with parameterized post-Friedmann approach}

\author{Lu Feng}
\affiliation{Department of Physics, College of Sciences, Northeastern University, Shenyang
110819, China}
\author{Yun-He Li}
\affiliation{Department of Physics, College of Sciences, Northeastern University, Shenyang
110819, China}
\author{Fei Yu}
\affiliation{College of Sciences, Shenyang Aerospace University, Shenyang 110136, China}
\author{Jing-Fei Zhang}
\affiliation{Department of Physics, College of Sciences, Northeastern University, Shenyang
110819, China}
\author{Xin Zhang\footnote{Corresponding author}}
\email{zhangxin@mail.neu.edu.cn} \affiliation{Department of Physics, College of Sciences,
Northeastern University, Shenyang 110819, China}
\affiliation{Center for High Energy Physics, Peking University, Beijing 100080, China}

\begin{abstract}
The model of holographic dark energy in which dark energy interacts with dark matter is investigated in this paper. In particular, we consider the interacting holographic dark energy model in the context of a perturbed universe, which was never investigated in the literature. To avoid the large-scale instability problem in the interacting dark energy cosmology, we employ the generalized version of the parameterized post-Friedmann approach to treat the dark energy perturbations in the model. We use the current observational data to constrain the model. Since the cosmological perturbations are considered in the model, we can then employ the redshift-space distortions (RSD) measurements to constrain the model, in addition to the use of the measurements of expansion history, which was either never done in the literature. We find that, for both the cases with $Q=\beta H\rho_{\rm c}$ and $Q=\beta H_0\rho_{\rm c}$, the interacting holographic dark energy model is more favored by the current data, compared to the holographic dark energy model without interaction. It is also found that, with the help of the RSD data, a positive coupling $\beta$ can be detected at the $2.95\sigma$ statistical significance for the case of $Q=\beta H_0\rho_{\rm c}$.

\end{abstract}
\maketitle

\section{Introduction}
\label{sec1}

Since the discovery of the cosmic acceleration~\cite{Riess:1998cb,Perlmutter:1998np}, the exploration of the nature of dark energy (DE) has become one of the most important issues in modern cosmology. The primary theoretical candidate for dark energy is the Einstein's cosmological constant $\Lambda$ that has a negative pressure, $p_\Lambda=-\rho_\Lambda$ (i.e., $w=-1$, with $w\equiv p/\rho$ being the equation-of-state parameter of dark energy). It should be mentioned that, the $\Lambda$CDM model, i.e., the cosmological model with $\Lambda$ and cold dark matter (CDM) can fit various cosmological observations fairly well by far. Although favored by the observations, the cosmological constant $\Lambda$ suffers from the fine-tuning and cosmic coincidence problems~\cite{Carroll:2000fy,Weinberg:2000yb}. To evade or alleviate these theoretical puzzles, numerous dynamical dark energy models have been proposed~\cite{Peebles:2002gy,Padmanabhan:2002ji,Copeland:2006wr,Frieman:2008sn,Li:2011sd,Li:2012dt,Bamba:2012cp}, such as the $w$CDM model, the $w_0$$w_a$CDM model, the holographic dark energy (HDE) model, and so forth. Among them, the HDE model~\cite{Li:2004rb} has attracted lots of attention, and has been studied widely~\cite{Huang:2004wt,Shen:2004ck,Zhang:2005hs,Zhang:2005yz,Wang:2004nqa,Huang:2004mx,Nojiri:2005pu,Zhang:2006qu,Chang:2005ph,Zhang:2006av,Setare:2006xu,Yi:2006bw,Zhang:2007sh,Zhang:2007es,Zhang:2007an,Ma:2007av,Li:2008zq,Ma:2007pd,Li:2009bn,Zhang:2009xj,delCampo:2011jp,Li:2013dha,Zhang:2015rha,Cui:2015oda,Landim:2015hqa,Wang:2016och,He:2016rvp}.

The HDE model~\cite{Li:2004rb} is a dynamical dark energy model based on a synthesis of the holographic principle of quantum gravity~\cite{tHooft:1993dmi,Susskind:1994vu} and effective quantum field theory. According to some consideration of the holographic principle, Cohen et al.~\cite{Cohen:1998zx} suggested that the total energy of a system with size $l$ should not exceed the mass of a black hole with the same size, leading to the inequality $l^3\rho_{\rm vac}\leqslant l M_{\rm pl}$, where $\rho_{\rm vac}$ is the vacuum energy density and $M_{\rm pl}$ is the reduced Planck mass. If the ultraviolet (UV) cutoff of the system is given, then the infrared (IR) length cutoff $L$ should be chosen by saturating the inequality, so that the holographic dark energy density is defined as
\begin{equation}
 \rho_{\rm de}=3c^2 M^2_{\rm pl} L^{-2},
\end{equation}
where $c$ is a dimensionless parameter characterizing some uncertainties in the effective quantum field theory. Li~\cite{Li:2004rb} pointed out that, in order to obtain a late-time accelerating universe, a reasonable option is to choose $L$ as the future event horizon of the universe, defined as
\begin{equation}\label{1.2}
L=a(t)\int_{t}^{\infty}\frac{dt'}{a(t')}=a\int_{a}^{\infty}\frac{da'}{Ha'^2},
\end{equation}
where $a$ is the scale factor of the universe, $H=\dot{a}/a$ is the Hubble parameter, and the dot denotes the derivative with respect to the cosmic time $t$. The HDE model has been proved to be a competitive and promising dark energy candidate~\cite{Li:2009jx,Xu:2016grp}, and even the cosmic coincidence problem can also be explained successfully in this model~\cite{Li:2004rb}. By far, various observational constraints on the HDE model all indicate that the parameter $c<1$, implying that the holographic dark energy would lead to a phantom universe with big-rip as its ultimate fate~\cite{Shen:2004ck,Huang:2004wt,Zhang:2005hs,Chang:2005ph,Yi:2006bw,Zhang:2007sh,Ma:2007pd,Li:2009bn}. Actually, by considering interaction between DE and DM in the HDE model, the big-rip problem can be effectively alleviated~\cite{Wang:2016lxa,Li:2009zs,Zhang:2012uu} (see also Ref.~\cite{Zhang:2009xj} for a solution by considering the extra dimension).


According to the present observations~\cite{Spergel:2003cb,Tegmark:2003ud,Abazajian:2004aja,Abazajian:2004it}, the current universe is dominated by two dark sectors, namely, DE and DM, where DE occupies about 70\% of the total energy while DM about 25\%. As enlightened by the quantum field theory, it is natural to consider that these two major components in the universe could have some direct non-gravitational interaction between them rather than evolve separately.
Actually, the interacting dark energy (IDE) models have been widely studied~\cite{Guo:2018gyo,Feng:2017usu,Guo:2017deu,Xia:2016vnp,Costa:2016tpb,Murgia:2016ccp,Yin:2015pqa,Duniya:2015nva,Geng:2015ara,Salvatelli:2014zta,Faraoni:2014vra,
Wang:2014iua,yang:2014vza,Yang:2014gza,Wang:2014oga,Zhang:2013zyn,Li:2013bya,Zhang:2013lea,Fu:2011ab,Li:2011ga,Xu:2011tsa,Li:2010ak,Li:2014eha,Li:2014cee,Li:2015vla,Guo:2017hea,Zhang:2017ize,Ma:2009uw,Li:2009zs,Zhang:2012uu,Feng:2016djj,Li:2017usw,Wang:2016lxa,Gomez-Valent:2014rxa,Sola:2015wwa,Sola:2016ecz,Sola:2017jbl,Sola:2017znb}. In particular, the interacting holographic dark energy (IHDE) model can not only alleviate the cosmic coincidence problem, but also help avoid the future big-rip singularity \cite{Wang:2016lxa,Li:2009zs,Zhang:2012uu}. See also Refs.~\cite{Ma:2009uw,Feng:2016djj,Li:2017usw} for further deep investigations on the IHDE model.

In our previous works~\cite{Feng:2016djj,Li:2017usw}, we have constrained the IHDE model by using the current observations. However, in these works, we only used the measurements of expansion history to constrain the model, and we did not consider the cosmological perturbations in the IHDE model. In fact, the cosmological perturbations have never been considered in the IHDE model in the literature. If one wishes to use the measurements of structure growth to constrain the IHDE model, the calculation of the cosmological perturbations in this model is a must. Under the circumstance of the nature of DE being unknown, the negative pressure of DE leads to that the sound speed of DE cannot be given in a general case, and the imposition of a rest-frame sound speed to DE by hand would usually lead to some instabilities for cosmological perturbations \cite{Zhao:2005vj,Valiviita:2008iv,He:2008si}. The instability in IDE models was found by Valiviita et al. \cite{Valiviita:2008iv} (see also Ref.~\cite{He:2008si}). To solve the instability problem in the IDE cosmology, Yun-He Li, Jing-Fei Zhang, and Xin Zhang generalized the parameterized post-Friedmann (PPF) framework to accommodate the IDE scenario \cite{Li:2014eha,Li:2014cee} (for the original PPF approach, see Refs.~\cite{Fang:2008sn,Hu:2008zd}). It has been shown that using the generalized version of the PPF framework the instability problem in the IDE cosmology could be successfully solved. Therefore, in the present work, we will apply the PPF approach in the IHDE model to consider the DE perturbations.


In this paper, for the first time we will consider the cosmological perturbations in the IHDE model. We will employ the generalized PPF approach to treat the DE perturbations in the IHDE model. Furthermore, we will constrain the IHDE model by using the current observations including the measurements of structure growth, which is also the first time. We wish to see whether a nonzero interaction can be detected by the current observations for the IHDE model in the case of considering cosmological perturbations.


The paper is organized as follows. In Sec. \ref{sec2}, we give a brief description of the PPF framework for the IDE scenario. In Sec.~\ref{sec3}, we present the analysis method and the observational data used in this work. In Sec. \ref{sec4}, we report the constraint results and discuss the relevant issues in detail. Conclusion is given in Sec. \ref{sec5}.

\section{A brief description of the PPF framework for interacting dark energy cosmology}
\label{sec2}

When considering a direct interaction between DE and CDM, the energy continuity equations for DE and CDM can be generally written as
\begin{align}
&\rho'_{\rm de} = -3\mathcal{H}(1+w)\rho_{\rm de}+ aQ_{\rm de},\label{eq1}\\
&\rho'_{\rm c} = -3\mathcal{H}\rho_{\rm c}+aQ_{\rm c},~~~~~~Q_{\rm de}=-Q_{\rm c}=Q,\label{eq2}
\end{align}
where $\rho_{\rm de}$ and $\rho_{\rm c}$ represent the energy densities of DE and CDM, respectively, a prime denotes the derivative with respect to the conformal time $\eta$, $w$ is the equation-of-state (EoS) parameter of DE, $\mathcal{H}=a'/a$ is the conformal Hubble parameter, and $Q$ denotes the energy transfer rate.


The EoS parameter of the holographic dark energy is given by~\cite{Li:2004rb}
\begin{equation}
w=-\frac{1}{3}-\frac{2}{3c\mathcal{H}}\sqrt{8\pi G\rho_{\rm de}a^2\over 3}.\label{w}
\end{equation}
Equations (\ref{eq1})--(\ref{w}), combined with the Friedmann equation, can easily determine the background evolution (expansion history) for the IHDE model.

For the form of $Q$, it is usually assumed to be proportional to the CDM density or the DE density, i.e., $Q=\beta H\rho_{\rm c}$ or $Q=\beta H\rho_{\rm de}$, where $\beta$ is the dimensionless coupling constant. However, there is another perspective that $Q$ should exclude the Hubble parameter $H$. This is because the local interactions ought not to rely on the overall expansion of the universe (see, e.g., Ref.~\cite{Valiviita:2008iv}). Thus, according to this perspective, another form of $Q$ is assumed to be, e.g., $Q=\beta H_0\rho_{\rm c}$ or $Q=\beta H_0\rho_{\rm de}$, where the appearance of the Hubble constant $H_0$ is only for a dimensional consideration.

There are several phenomenological forms of $Q$ often discussed in the literature. In this work, we only consider two cases with $Q=\beta H\rho_{\rm c}$ (denoted as $Q_1$) and $Q=\beta H_0\rho_{\rm c}$ (denoted as $Q_2$).
According to Eqs. (\ref{eq1}) and (\ref{eq2}), $\beta>0$ means the decay of CDM into DE, $\beta<0$ means the decay of DE into CDM, and obviously $\beta=0$ means no interaction.
For convenience, in this paper, the IHDE models with $Q_1$ and $Q_2$ are denoted as the IHDE1 model and the IHDE2 model, respectively.

In the covariant formalism, the conservation laws for DE and CDM can be expressed as
\begin{equation}
\label{eqn:energyexchange} \nabla_\nu T^{\mu\nu}_I = Q^\mu_I, \quad\quad
 \sum_I Q^\mu_I = 0,
\end{equation}
where $T^{\mu\nu}_I$ is the energy-momentum tensor for $I=\rm de$ and $\rm c$, and $Q^\mu_I$ is the energy-momentum transfer vector. Here we choose $Q^\mu_{\rm de}=-Q^\mu_{\rm c}=Qu^\mu_{\rm c}$, where $u^\mu_{\rm c}$ is the four-velocity of CDM. The energy-momentum transfer vector can be split into two parts as
\begin{equation}
Q^I_\mu  = a\big( -Q_I(1+AY) - \delta Q_IY,\,[ f_I+ Q_I (v-B)]Y_i\big),\label{eq:Qenergy}
\end{equation}
where $\delta Q_I$ is the energy transfer perturbation and $f_I$ is the momentum transfer potential of the $I$ fluid.  $A$ and $B$ are the scalar metric perturbations. $Y$ and $Y_i$ are the eigenfunctions of the Laplace operator and its covariant derivative.

Equations~(\ref{eqn:energyexchange}) and (\ref{eq:Qenergy}) then lead to the following conservation equations for the $I$ fluid in the IDE scenario,
\begin{equation}
 {\delta\rho_I'}
	+  3\mathcal{H}({\delta \rho_I}+ {\delta p_I})+(\rho_I+p_I)(k{v}_I + 3 H_L')=a(\delta Q_I+AQ_I),\label{eqn:conservation1}
\end{equation}	
\begin{eqnarray}	
[(\rho_I + p_I)({{v_I}-{B}})]'+4\mathcal{H}(\rho_I + p_I)({{v_I}-{B}})-k{ \delta p_I }\nonumber \\
+ {2 \over 3}k\ck p_I {\Pi_I} - k(\rho_I+ p_I) {A}=a[Q_I(v-B)+f_I],\label{eqn:conservation2}
\end{eqnarray}
where $\delta\rho_I$ is energy density perturbation, $\delta p_I$ is isotropic pressure perturbation, $v_I$ is velocity perturbation, $\Pi_I$ is anisotropic stress perturbation, and $c_K = 1-3K/k^2$ with $K$ being the spatial curvature.

In the conventional way~\cite{Valiviita:2008iv}, DE is treated as a nonadiabatic fluid and the calculation of $\delta p_{\rm de}$ is in terms of the adiabatic sound speed and the rest-frame sound speed, therefore in the IDE scenario the large-scale instability will occasionally occur. To avoid the instability problem in the IDE cosmology, we treat the DE perturbations by employing the generalized PPF scheme~\cite{Li:2014eha}. In the following, we give a brief description of the PPF method for the IDE scenario. Note that, to avoid unnecessary confusion, we use the new symbols, i.e., $\zeta\equiv H_L$, $\xi\equiv A$, $\rho\Delta\equiv\delta\rho$, $\Delta p\equiv\delta p$, $V\equiv v$, and $\Delta Q_I\equiv\delta Q_I$, to denote the corresponding quantities of the comoving gauge, except the two gauge-independent quantities $\Pi$ and $f_I$.

On large scales, a direct relationship between $V_{\rm de} - V_T$ and $V_T$ is established, where the subscript ``T'' denotes the total matter except DE. This relationship can be parametrized by a function $f_\zeta(a)$ as \cite{Hu:2008zd,Fang:2008sn}
\begin{equation}
\lim_{k_H \ll 1}
 {4\pi G a^2\over \mathcal{H}^2} (\rho_{\rm de} + p_{\rm de}) {V_{\rm de} - V_T \over k_H}
= - {1 \over 3} \ck  f_\zeta(a) k_H V_T,\label{eq:DEcondition}
\end{equation}
where $k_H=k/\mathcal{H}$. Combining this condition and the Einstein equations, we get the equation of motion for the curvature perturbation $\zeta$ on the large scales,
\begin{align}
\lim_{k_H \ll 1} \zeta'  = \mathcal{H}\xi - {K \over k} V_T +{1 \over 3} \ck  f_\zeta(a) k V_T.
\label{eqn:zetaprimesh}
\end{align}
On small scales, the Poisson equation is used to describe the evolution of the curvature perturbation, $\Phi=4\pi G a^2\Delta_T \rho_T/( k^2\ck)$, with $\Phi=\zeta+V_T/k_H$. In order to make these two limits compatible, one can introduce a dynamical function $\Gamma$ so that
\begin{equation}
\Phi+\Gamma = {4\pi Ga^2
\over  k^2\ck} \Delta_T \rho_T
\label{eqn:modpoiss}
\end{equation}
is satisfied on all scales.

Compared with the small-scale Poisson equation, Eq.~(\ref{eqn:modpoiss}) gives $\Gamma\rightarrow0$ at $k_H\gg1$. By taking the derivative of Eq.~(\ref{eqn:modpoiss}) and using the conservation equations and the Einstein equations [Eqs.~(\ref{eqn:conservation1}), (\ref{eqn:conservation2}), and (\ref{eqn:zetaprimesh})], one derives the equation of motion for $\Gamma$ on the large scales,
\begin{equation}\label{eq:gammadot}
\lim_{k_H \ll 1} \Gamma'  = S -\mathcal{H}\Gamma,
\end{equation}
with
\begin{align}
S&={4\pi Ga^2
\over k^2 } \Big\{[(\rho_{\rm de}+p_{\rm de})-f_{\zeta}(\rho_T+p_T)]kV_T \nonumber\\
&\quad+{3a\over k_Hc_K}[Q_{\rm c}(V-V_T)+f_{\rm c}]+\frac{a}{c_K}(\Delta Q_{\rm c}+\xi Q_{\rm c})\Big\},\nonumber
\end{align}
where $\xi$ can be obtained from Eq.~(\ref{eqn:conservation2}),
\begin{equation}
\xi =  -{\Delta p_T - {2\over 3}\tppi+{a\over k}[Q_{\rm c}(V-V_T)+f_{\rm c}] \over \rho_T + p_T}.
\label{eqn:xieom}
\end{equation}

Using a parameter $c_\Gamma$ that gives a transition scale in terms of the Hubble scale, under which DE is smooth enough, we can take the equation of motion for $\Gamma$ on all scales to be \cite{Hu:2008zd,Fang:2008sn}
\begin{equation}
(1 + c_\Gamma^2 k_H^2) [\Gamma' +\mathcal{H} \Gamma + c_\Gamma^2 k_H^2 \mathcal{H}\Gamma] = S.
\label{eqn:gammaeom}
\end{equation}

Note that, in the equation of motion for $\Gamma$, all of the perturbation quantities are those of matter excluding DE. Therefore, although one has no any knowledge of the DE perturbations, the differential equation (\ref{eqn:gammaeom}) can be solved. Once the evolution of $\Gamma$ is obtained, we can directly get the energy density and velocity perturbations,
\begin{align}
&\rho_{\rm de}\Delta_{\rm de} =- 3(\rho_{\rm de}+p_{\rm de}) {V_{\rm de}-V_{T}\over k_{H} }-{k^{2}\ck \over 4\pi G a^{2}} \Gamma,\\ \label{eqn:ppffluid}
& V_{\rm de}-V_{T} ={-k \over 4\pi Ga^2 (\rho_{\rm de} + p_{\rm de}) F} \nonumber \\
&\quad\quad\quad\times\left[ S - \Gamma' - \mathcal{H}\Gamma + f_{\zeta}{4\pi Ga^2 (\rho_{T}+p_{T}) \over k}V_{T}
\right],
\end{align}
with $F = 1 +  12 \pi G a^2 (\rho_T + p_T)/( k^2 \ck)$.

We apply this generalized version of the PPF method to the IHDE model in the following numerical calculations. The PPF method avoids the use of pressure perturbation defined by sound speed and can help us to probe the whole parameter space of the IHDE model. For more information about the PPF scheme for IDE cosmology, we refer the reader to Refs.~\cite{Li:2014eha,Li:2014cee}.

\section{Method and data}\label{sec3}

For the IHDE model, there are eight base parameters, which are the physical baryon density $\Omega_{\rm b}h^2$, the physical cold dark matter density $\Omega_{\rm c}h^2$, the ratio of the sound horizon and angular diameter distance at the time of last-scattering $\theta_{\rm MC}$, the HDE model parameter $c$, the coupling constant $\beta$, the reionization optical depth $\tau$, and the amplitude $A_s$ and the tilt $n_s$ of the primordial scalar fluctuations.
To infer the posterior probability distributions of parameters, we use the public Markov-chain Monte Carlo (MCMC) package CosmoMC~\cite{Lewis:2002ah} to perform the calculations. In addition, we use the PPF package~\cite{Li:2014eha,Li:2014cee,Li:2015vla,Guo:2017hea,Zhang:2017ize} for the IHDE model to handle the perturbations of dark energy.

In this paper, we will also make a comparison for the relevant models. Note that, these models (HDE and IHDE) have different numbers of parameters, i.e., the HDE model has $7$ parameters while the IHDE model has 8 parameters. Thus, when we perform a comparison for these models, from a statistical point of view, the simple comparison of $\chi^2$ is obviously unfair. Therefore, in this work, we simply adopt the Akaike information criterion (AIC) \cite{AIC1974} for the model comparison. By definition, we have ${\rm AIC}=\chi^2_{\rm{min}}+2k$, where $k$ is the number of parameters. The model with a lower value of AIC is more favored by data.

The observations we use in this work are comprised of the Planck cosmic microwave background (CMB) data, the baryon acoustic oscillations (BAO) data, the type Ia supernovae (SN) data, the Hubble constant ($H_0$) direct measurement data, and the RSD data.
\begin{itemize}
\item  The Planck data: We use the Planck CMB full temperature and polarization power spectra data, including the TT, TE, EE and lowP data, released in 2015 \cite{Aghanim:2015xee}.
\item  The BAO data: We use the recent BAO measurement from the Six-Degree-Field Galaxy Survey (6dFGS) at $z_{\rm eff}=0.106$ \cite{Beutler:2011hx}, the Main Galaxy Sample of Data Release 7 of Sloan Digital Sky Survey (SDSS-MGS) at $z_{\rm eff}=0.15$ \cite{Ross:2014qpa}, and the CMASS and LOWZ samples of Data Release 12 of Baryon Oscillation Spectroscopic Survey (BOSS) at $z_{\rm eff}=0.57$ and $z_{\rm eff}=0.32$ \cite{Cuesta:2015mqa}, respectively.
\item  The SN data: We use the Joint Light-curve Analysis (JLA) sample of the SN observation, compiled from the SNLS, SDSS, and several samples of low-redshift SN data \cite{Betoule:2014frx}.
\item  The $H_0$ data: We use the latest result of the Hubble constant direct measurement, given by Riess et al.~\cite{Riess:2016jrr}, with the measurement value $H_0=73.00{\pm1.75}~{\rm km}~{\rm s}^{-1}~{\rm Mpc}^{-1}$.
\item  The RSD data: We employ the RSD measurements from VIPERS ($z=0.80$)~\cite{RSDvipers}, WiggleZ ($z=0.22, 0.41, 0.60$ and $0.78$)~\cite{RSDwigglez}, BOSS CMASS DR12 ($z=0.57$) and LOWZ DR12 ($z=0.32$)~\cite{Gil-Marin:2016wya}, SDSS LRG DR7 ($z=0.25$ and $z=0.37$)~\cite{RSDsdss7}, 2dFGS ($z=0.17$)~\cite{RSD2dF} and 6dFGS ($z=0.067$)~\cite{RSD6dF}.
\end{itemize}

For simplicity, we use ``BSH" to denote the combination of BAO, SN and $H_0$.
In our analysis, we use two sets of data combination: (i) Planck+BSH and (ii) Planck+BSH+RSD. In the next section, we will report and discuss the fitting results with these two data sets.



\begin{table*}\small
\setlength\tabcolsep{1.5pt}
\renewcommand{\arraystretch}{1.5}
\caption{\label{taball}Fitting results for the HDE ($Q=0$), IHDE1 ($Q=\beta H\rho_{\rm c}$), and IHDE2 ($Q=\beta H_0\rho_{\rm c}$) models from the data combinations Planck+BSH and Planck+BSH+RSD.}
\centering
\begin{tabular}{cccccccccccc}
\hline \multicolumn{1}{c}{Model} &&\multicolumn{2}{c}{$Q=0$}&&\multicolumn{2}{c}{$Q=\beta H\rho_{\rm c}$}&&\multicolumn{2}{c}{$Q=\beta H_0\rho_{\rm c}$}&\\
           \cline{1-1}\cline{3-4}\cline{6-7}\cline{9-10}

 Data  && Planck+BSH &Planck+BSH+RSD&& Planck+BSH &Planck+BSH+RSD&&Planck+BSH &Planck+BSH+RSD\\
\hline

$\Omega_{\rm m}$     &&$0.289\pm0.008$
               &$0.295\pm0.008$
               &&$0.284\pm0.008$
               &$0.289\pm0.008$
               &&$0.247^{+0.017}_{-0.020}$
               &$0.241^{+0.017}_{-0.019}$\\

$\sigma_8$     &&$0.836\pm0.017$
               &$0.803\pm0.013$
               &&$0.838\pm0.016$
               &$0.807\pm0.013$
               &&$0.840\pm0.017$
               &$0.814\pm0.014$\\

$H_0\,[{\rm km}/{\rm s}/{\rm Mpc}]$          &&$69.74^{+0.94}_{-0.93}$
               &$68.73^{+0.86}_{-0.85}$
               &&$69.84^{+0.92}_{-0.93}$
               &$68.90\pm0.88$
               &&$69.91^{+0.94}_{-0.97}$
               &$69.03\pm0.88$\\

$\beta$        &&...
               &...
               &&$0.0034^{+0.0016}_{-0.0018}$
               &$0.0044^{+0.0017}_{-0.0019}$
               &&$0.207^{+0.091}_{-0.093}$
               &$0.271^{+0.090}_{-0.092}$\\

$c$            &&$0.600^{+0.027}_{-0.031}$
               &$0.644\pm0.029$
               &&$0.648^{+0.037}_{-0.045}$
               &$0.708^{+0.039}_{-0.048}$
               &&$0.770^{+0.080}_{-0.102}$
               &$0.873^{+0.084}_{-0.102}$\\
\hline
$\chi^2_{\rm min}$     &&13683.030
                       &13702.878
                       &&13681.034
                       &13695.808
                       &&13680.612
                       &13695.128\\
\hline
\end{tabular}

\end{table*}

\begin{figure*}[!htp]
\includegraphics[scale=0.7]{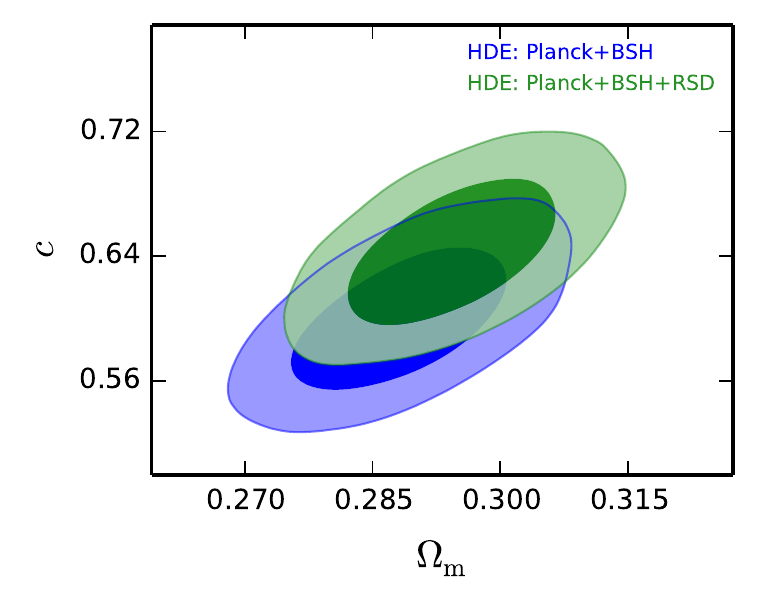}
\includegraphics[scale=0.7]{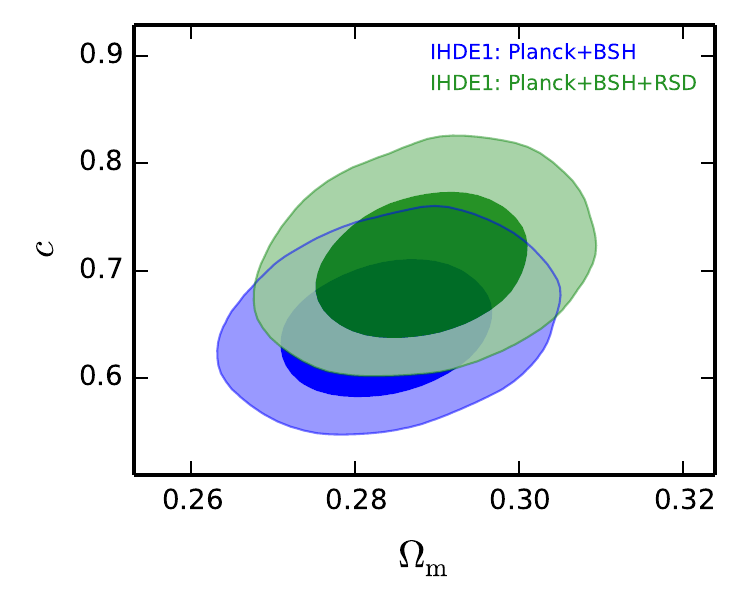}
\includegraphics[scale=0.7]{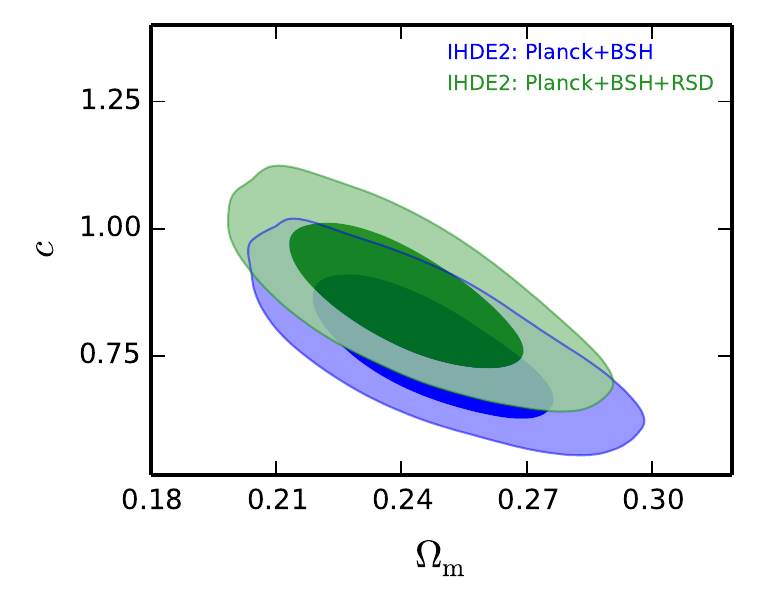}
\centering
\caption{\label{allomc}The two-dimensional marginalized contours (68.3\% and 95.4\% confidence level) in the $\Omega_{\rm m}$--$c$ plane for the HDE ($Q=0$), IHDE1 ($Q=\beta H\rho_{\rm c}$), and IHDE2 ($Q=\beta H_0\rho_{\rm c}$) models by using the Planck+BSH and Planck+BSH+RSD data combinations.}
\end{figure*}

\begin{figure*}[!htp]
\includegraphics[scale=0.7]{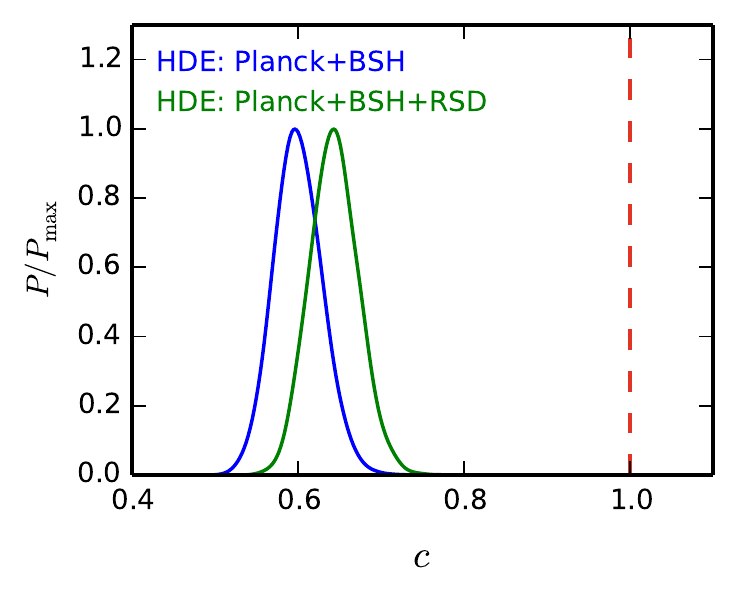}
\includegraphics[scale=0.7]{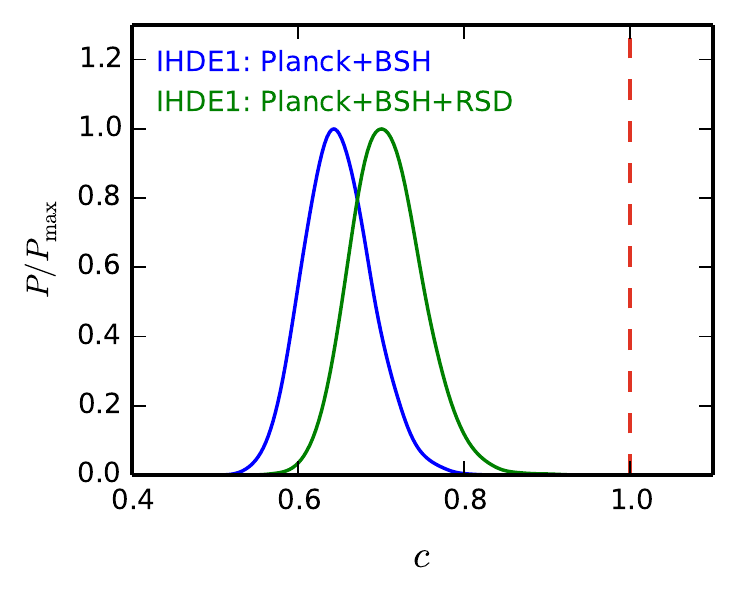}
\includegraphics[scale=0.7]{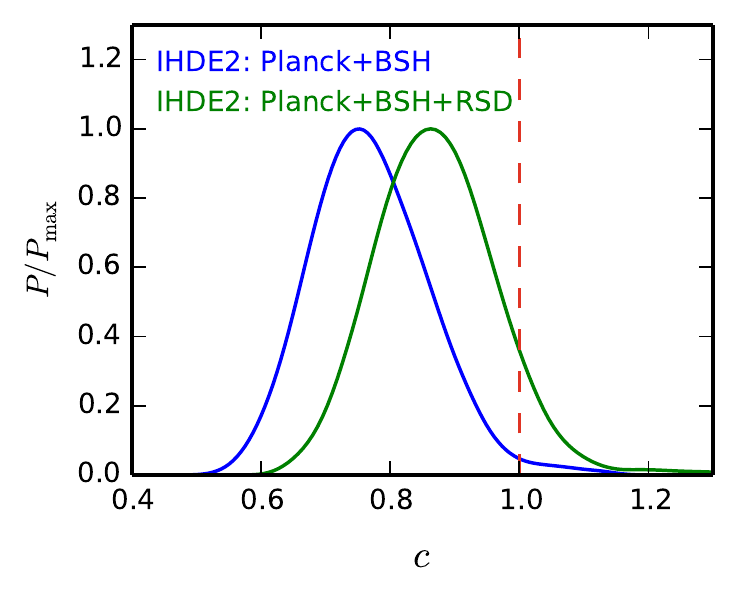}
\centering
\caption{\label{likechh0}The one-dimensional posterior distributions of $c$ for the HDE ($Q=0$), IHDE1 ($Q=\beta H\rho_{\rm c}$), and IHDE2 ($Q=\beta H_0\rho_{\rm c}$) models by using the Planck+BSH and Planck+BSH+RSD data combinations.}
\end{figure*}

\begin{figure}[ht!]
\begin{center}
\includegraphics[width=8.0cm]{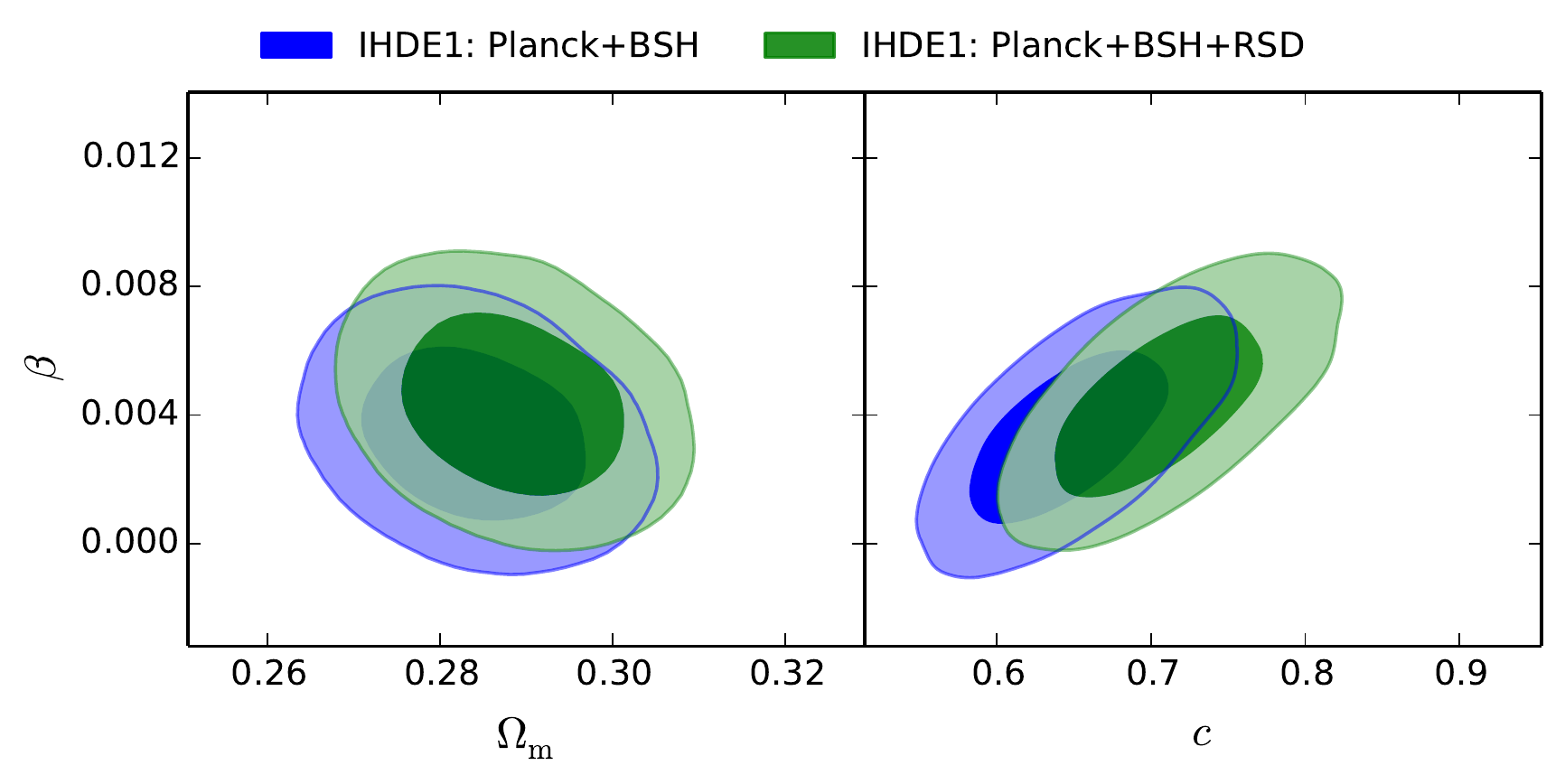}
\includegraphics[width=8.0cm]{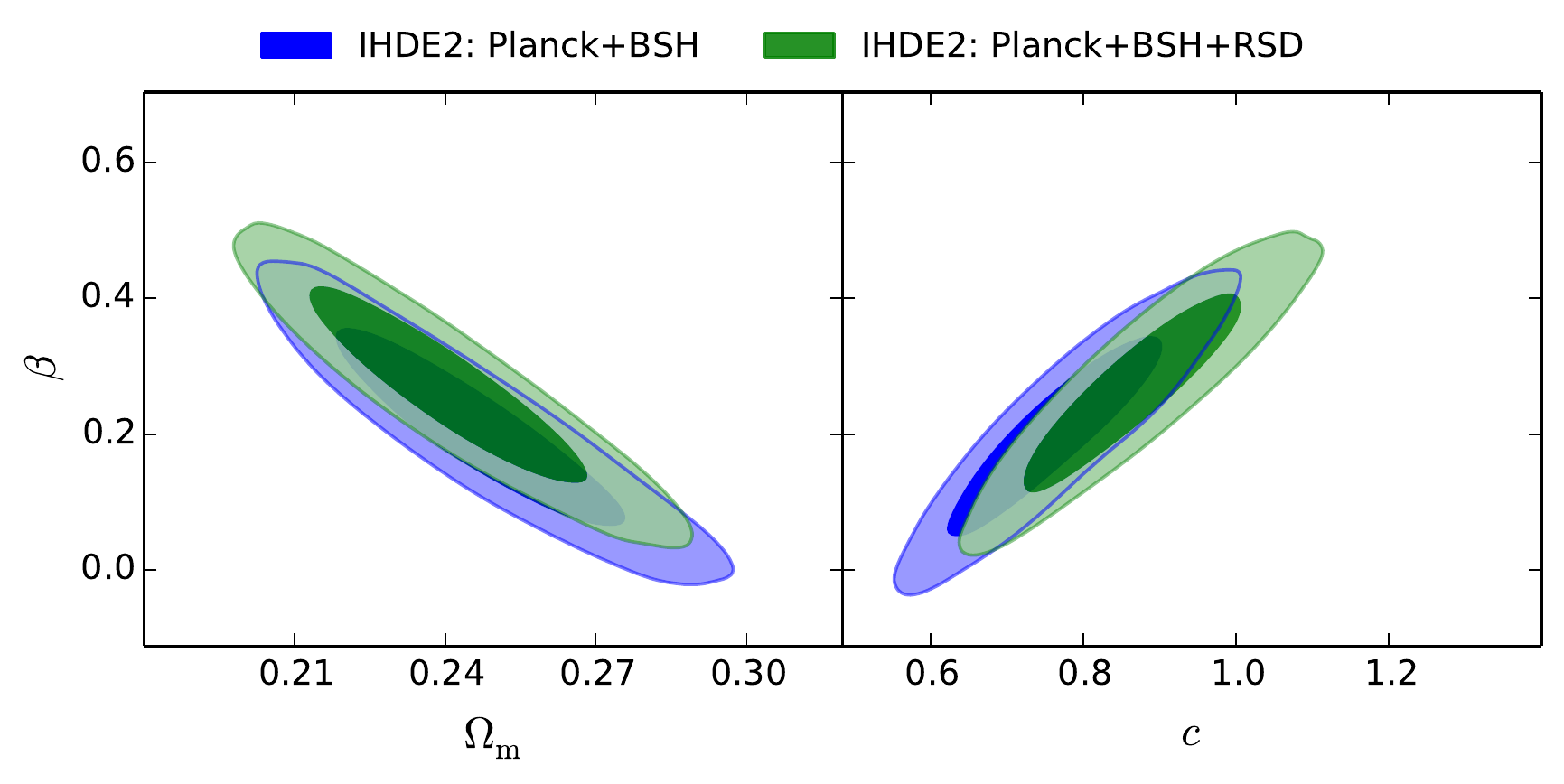}
\end{center}
\caption{\label{allhh0}The two-dimensional marginalized contours (68.3\% and 95.4\% confidence level) in the $\Omega_{\rm m}$--$\beta$ and $c$--$\beta$ planes for the IHDE1 ($Q=\beta H\rho_{\rm c}$) and IHDE2 ($Q=\beta H_0\rho_{\rm c}$) models by using the Planck+BSH and Planck+BSH+RSD data combinations.}
\end{figure}

\begin{figure}[ht!]
\begin{center}
\includegraphics[width=6cm]{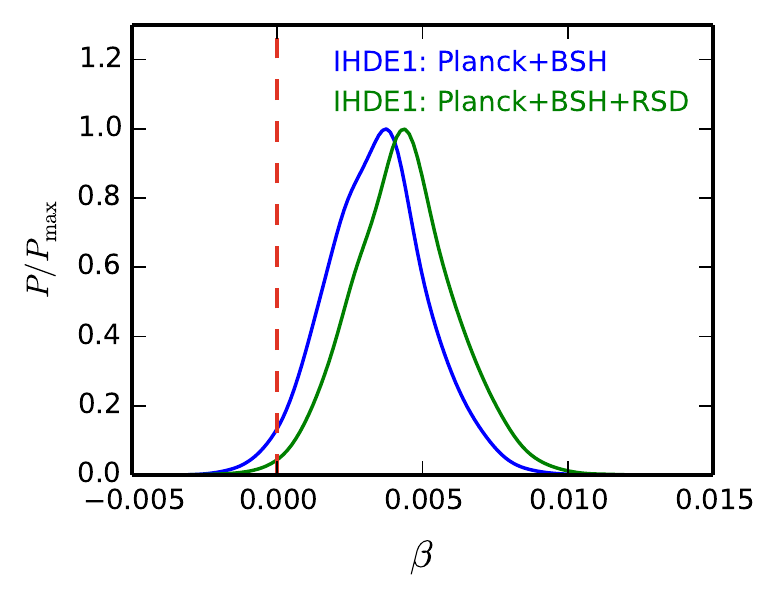}
\includegraphics[width=6cm]{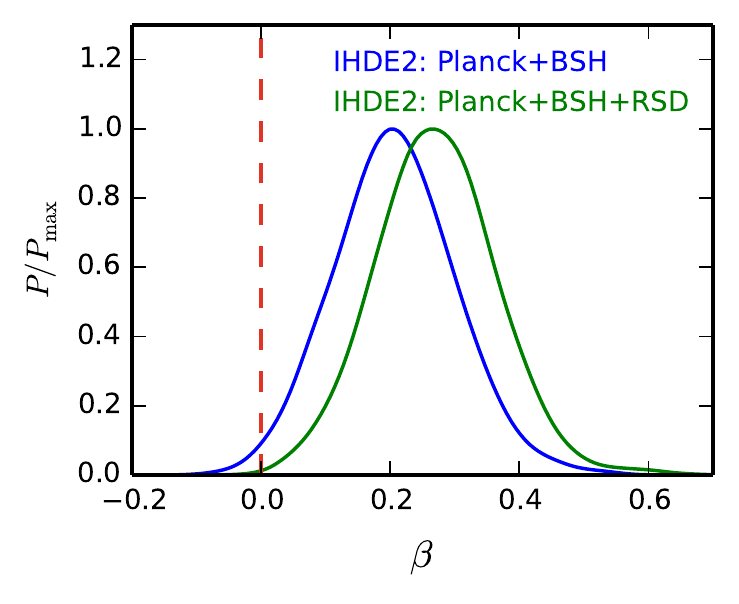}
\centering
\caption{\label{likebhh0}The one-dimensional posterior distributions of $\beta$ for the IHDE1 ($Q=\beta H\rho_{\rm c}$) and IHDE2 ($Q=\beta H_0\rho_{\rm c}$) models by using the Planck+BSH and Planck+BSH+RSD data combinations. }
\end{center}
\end{figure}

\section{Results and discussion}\label{sec4}

In this section, we report the fitting results of the IHDE models and discuss the implications of them. We use the Planck+BSH and Planck+BSH+RSD data combinations to constrain the IHDE models. For a comprehensive comparison, the fitting results for the HDE model without interaction from the same two data sets are also shown. The fitting results are given in Table~\ref{taball}, where the $\pm1\sigma$ errors are quoted.

In Table~\ref{taball}, the values of $\chi^2_{\rm{min}}$ are summarized for the two data combinations, Planck+BSH and Planck+BSH+RSD.
In the case of Planck+BSH constraints, for the HDE model we have $\chi^2_{\rm{min}}=13683.030$. The IHDE model (with one more parameter than HDE) yields a decrease for the value of $\chi^2_{\rm{min}}$, compared with HDE, by $\Delta\chi^2=-1.996$ (IHDE1) and $\Delta\chi^2=-2.418$ (IHDE2). This indicates that the IHDE models can slightly improve the fit.
In the case of Planck+BSH+RSD constraints, the IHDE1 model leads to a decrease of $\Delta\chi^2=-7.070$ and the IHDE2 model leads to a decrease of $\Delta\chi^2=-7.750$.
This indicates that, with the help of the RSD data, the IHDE models can evidently improve the fit.
Moreover, we further make a model selection by using the information criterion. We have $\Delta {\rm AIC}=0.004$ for the IHDE1 model and
$\Delta {\rm AIC}=-0.418$ for the IHDE2 model with the Planck+BSH data set. Clearly, the IHDE1 model is slightly worse than the HDE model and the IHDE2 model is slightly better than the HDE model.
Using the Planck+BSH+RSD data set, we have $\Delta {\rm AIC}=-5.070$ for the IHDE1 model and $\Delta {\rm AIC}=-5.750$ for the IHDE2 model. Thus, after considering the RSD data, we find that the IHDE models are evidently better
than the HDE model, and the IHDE2 ($Q=\beta H_0\rho_{\rm c}$) model is slightly better than the IHDE1 ($Q=\beta H\rho_{\rm c}$) model from the statistical point of view.

The fitting results of the HDE model and the IHDE models are shown in Table~\ref{taball}. It can be clearly seen that the values of $c<1$ are obtained in both the HDE and IHDE models.
For the HDE model, we obtain $c=0.600^{+0.027}_{-0.031}$ by using the Planck+BSH data and $c=0.644\pm0.029$ by using the Planck+BSH+RSD data, indicating that $c<1$ at $14.8\sigma$ and $12.3\sigma$, respectively.
For the IHDE1 model, we obtain $c=0.648^{+0.037}_{-0.045}$ by using the Planck+BSH data and $c=0.708^{+0.039}_{-0.048}$ by using the Planck+BSH+RSD data, indicating that $c<1$ at $9.5\sigma$ and $7.5\sigma$, respectively.
For the IHDE2 model, we obtain $c=0.770^{+0.080}_{-0.102}$ by using the Planck+BSH data and $c=0.873^{+0.084}_{-0.102}$ by using the Planck+BSH+RSD data, indicating that $c<1$ at $2.9\sigma$ and $1.5\sigma$, respectively.
Obviously, if the interaction is considered in the scenario of holographic dark energy, the statistical significance of $c<1$ will be decreased.
In addition, we find that, when the RSD data are considered in these models, a relatively larger $c$ will be obtained.
Therefore, from this analysis, it is found that (i) the RSD data could significantly influence the constraints on $c$ and (ii) considering interaction between DE and DM in the scenario of holographic dark energy could largely decrease the risk of the future big-rip singularity. In particular, in the IHDE2 ($Q=\beta H_0\rho_{\rm c}$) model the risk of big rip would be substantially decreased.
In Figs.~\ref{allomc} and~\ref{likechh0}, we show the posterior distribution contours in the $\Omega_{\rm m}$--$c$ plane and the one-dimensional posterior distributions for $c$, respectively, for the HDE model and the IHDE models.

In the case of Planck+BSH constraints, from Table~\ref{taball}, we have $\beta=0.0034^{+0.0016}_{-0.0018}$ for the IHDE1 model and $\beta=0.207^{+0.091}_{-0.093}$ for the IHDE2 model, respectively. Evidently, for both of the two IHDE models, we find that the Planck+BSH data prefer a positive value of $\beta$ at $1.89\sigma$ level (IHDE1) and $2.23\sigma$ level (IHDE2), indicating that cold dark matter decays into dark energy. Thus, a null interaction is excluded at about $2\sigma$ level in the IHDE models by using the Planck+BSH data.
But, for the IHDE models investigated in Refs.~\cite{Feng:2016djj,Li:2017usw}, in which the consideration of the cosmological perturbations was absent,
it was found that in the $Q=\beta H\rho_{\rm c}$ and $Q=\beta H_0\rho_{\rm c}$ models
a null interaction can be excluded only at less than $1.50\sigma$ significance with the data combination of CMB+BAO+SN+$H_0$ (note that, for the observations, they use the ``CMB distance priors" from the 2015 release of Planck~\cite{Ade:2015rim}, the BAO data from Refs.~\cite{Beutler:2011hx,Ross:2014qpa,Anderson:2013zyy}, and the $H_0$ data from Ref.~\cite{Efstathiou:2013via}). From the above analysis, we find that the cosmological perturbations could have a significant impact on the measurement of the coupling constant $\beta$.

Besides, the dark energy properties could also impact the cosmological constraints on the coupling constant $\beta$. It is of great interest to see how different DE models affect the constraints on $\beta$.
The cosmological constraints on $\beta$ in a scenario of vacuum energy interacting with cold dark matter (I$\Lambda$CDM) have been discussed in Refs.~\cite{Feng:2017usu,Guo:2017hea}, where it was found that $\beta=0.0021\pm0.0011$ for the $Q=\beta H\rho_{\rm c}$ model by using the Planck+BSH data combination, the same data set as used in this paper, indicating that $\beta>0$ at $1.91\sigma$ level.
Thus, we find that the dark energy properties could influence the constraint limits of the coupling constant $\beta$. Compared with the I$\Lambda$CDM model, we find that in the IHDE model the central value of $\beta$ is increased and the error range of $\beta$ is amplified.

Furthermore, we wish to see how the inclusion of the RSD data help constrain the coupling constant $\beta$.
The fit results are also shown in Table~\ref{taball}.
By using the Planck+BSH+RSD data, we obtain $\beta=0.0044^{+0.0017}_{-0.0019}$ for the IHDE1 ($Q=\beta H\rho_{\rm c}$) model and $\beta=0.271^{+0.090}_{-0.092}$ for the IHDE2 ($Q=\beta H_0\rho_{\rm c}$) model, indicating that $\beta>0$ at $2.32\sigma$ and $2.95\sigma$, respectively. Obviously, we can see that the inclusion of the RSD data favors a relatively larger $\beta$ for both of the two IHDE models.
It is of great interest to find that the detection of $\beta>0$ turns out to be at more than $2\sigma$ level in the IHDE models.
In particular, for the IHDE2 model, $\beta>0$ is favored at $2.95\sigma$ by using the Planck+BSH+RSD data.
Thus, a null interaction is excluded at about $3\sigma$ level, showing that the interaction is preferred with the help of the RSD data (see also Ref.~\cite{Salvatelli:2014zta} for case of the I$\Lambda$CDM cosmology, in which it was found that a null interaction is excluded at $3\sigma$ level with help of the RSD data).
In addition, Fig.~\ref{allhh0} shows that $\beta$ is positively correlated with $c$, namely, a larger $\beta$ leads to a larger $c$, thus the risk of becoming a phantom for HDE can be decreased as discussed above.
To show apparently the effect of the RSD data on the constraints on $\beta$, the one-dimensional posterior distributions of $\beta$ for the IHDE models are plotted in Fig.~\ref{likebhh0}.

\section{Conclusion}\label{sec5}

In this paper, we have studied two interacting holographic dark energy models with the energy transfer forms $Q=\beta H\rho_{\rm c}$ and $Q=\beta H_0\rho_{\rm c}$, respectively. We adopt the PPF approach to calculate the perturbations of dark energy. The current observational data used here include the Planck 2015 CMB temperature and polarization data, the BAO data, the JLA compilation of SN data, the $H_0$ direct measurement, and the RSD data.

We find that the current observations slightly favor the $Q=\beta H_0\rho_{\rm c}$ model over the $Q=\beta H\rho_{\rm c}$ model, and both of the two IHDE models fit the current observations better than the HDE model does.
We also find that the statistical significance of $c<1$ will decrease, when considering the interaction between dark energy and dark matter.
In particular, for the IHDE2 model, we have $c=0.873^{+0.084}_{-0.102}$ with the inclusion of the RSD data, indicating $c<1$ only at the $1.5\sigma$ level.
Therefore, with the help of the interaction, the risk of becoming a phantom for holographic dark energy is decreased.

In addition, by using the Planck+BSH data, we obtain $\beta=0.0034^{+0.0016}_{-0.0018}$ for the $Q=\beta H\rho_{\rm c}$ model and $\beta=0.207^{+0.091}_{-0.093}$ for the $Q=\beta H_0\rho_{\rm c}$ model. This indicates that a positive $\beta$ is favored in the IHDE models. By using the Planck+BSH+RSD data, we obtain $\beta=0.0044^{+0.0017}_{-0.0019}$ for the $Q=\beta H\rho_{\rm c}$ model and $\beta=0.271^{+0.090}_{-0.092}$ for the $Q=\beta H_0\rho_{\rm c}$ model. We find that the current RSD data favor a larger interaction rate for the models studied, and the coupling $\beta>0$ can be detected at more than $2\sigma$ level.
For the $Q=\beta H\rho_{\rm c}$ model, $\beta>0$ is favored at $2.23\sigma$ significance and for the $Q=\beta H_0\rho_{\rm c}$ model, $\beta>0$ can be detected at $2.95\sigma$ significance.
Thus the $Q=\beta H_0\rho_{\rm c}$ model deserves further deeper investigations in the next step.

\begin{acknowledgments}
This work was supported by the National Natural Science Foundation of China (Grant Nos.~11522540, 11690021, 11875102, 11805031, and 11835009), the Top-Notch Young Talents Program of China, the Provincial Department of Education of Liaoning (Grant No.~L2012087), and the Fundamental Research Funds for the Central Universities (Grant No.~N170503009).

\end{acknowledgments}

\end{document}